\documentclass[letterpaper, 10 pt, conference]{ieeeconf}  

\usepackage{amsmath, amssymb}
\usepackage{graphicx}
\usepackage{algorithm}
\usepackage{algorithmic}
\usepackage{booktabs}
\usepackage{float}
\usepackage[dvipsnames]{xcolor}

\usepackage[left=2.5cm, right=1.3cm, top=2.5cm, bottom=2.5cm]{geometry}
\usepackage{amsmath}
\usepackage{amssymb}

\usepackage{amsthm}

\usepackage{comment}
\usepackage{dsfont} 

\newtheorem{assumption}{Assumption}
\newtheorem{definition}{Definition}
\newtheorem{remark}{Remark}

\newtheoremstyle{plain}
  {} 
  {} 
  {\itshape} 
  {} 
  {\bfseries} 
  {.} 
  {.5em} 
  {} 
\theoremstyle{plain}

\newtheorem{theorem}{Theorem}

\IEEEoverridecommandlockouts                              

\title{\LARGE \bf
Robust Offset-free Kernelized Data-Driven Predictive Control for Nonlinear Systems
 }

\author{ Mahmood Mazare, Hossein Ramezani 
\thanks{The authors are with the Department of Mechanical and Electrical Engineering, University of Southern Denmark, Denmark.
        {\tt\small \{mazare\}@sdu.dk}}%
}

\begin{document}

\maketitle
\thispagestyle{empty}
\pagestyle{empty}


\begin{abstract}
This paper presents a Kernelized Data-Driven Predictive Control (KDPC) scheme for robust, offset-free tracking of nonlinear systems. To overcome the computational burden of direct data-driven methods, we employ a hybrid framework that learns the nonlinear dynamics in a Reproducing Kernel Hilbert Space (RKHS) via joint ridge regression. A key contribution is the derivation of an analytical linearization of the kernel map, which renders the control problem a strictly convex Quadratic Program (QP) for efficient real-time implementation. We provide rigorous guarantees for recursive feasibility using terminal ingredients and establish Input-to-State Stability (ISS) with respect to the kernel approximation error. The efficacy of the proposed KDPC is demonstrated on a Van der Pol oscillator, showing disturbance rejection and offset-free performance compared to standard model-based benchmarks.
\end{abstract}

\section{Introduction}

Data-Driven Predictive Control (DPC) has gained significant traction over the last decade, challenging traditional model-based control paradigms \cite{xiong2025data,krishnan2021direct}. The key advantage of DPC is its ability to bypass the complex, time-intensive task of first-principles modeling; instead, it achieves high-performance control by leveraging input-output data directly \cite{huang2023robust}. Methodologically, DPC strategies are typically bifurcated into two main classes: direct and indirect methods \cite{hou2013model,markovsky2021behavioral,huang2023robust}.

Indirect methods, such as Subspace Predictive Control (SPC) \cite{favoreel1999spc}, first identify a predictive model from data, which is then used within a standard Model Predictive Control (MPC) optimization. In contrast, direct DPC approaches, most notably Data-Enabled Predictive Control (DeePC) \cite{coulson2019data}, are built upon Willems' fundamental lemma \cite{willems2005note}. These methods incorporate historical data directly into the control problem by introducing a decision variable, $\mathbf{g}(k)$, which represents a linear combination of past system trajectories. While this direct formulation is elegant, the dimension of $\mathbf{g}(k)$ scales with the amount of data, often leading to large-scale optimization problems that are computationally intensive to solve in real-time.

To address this challenge, hybrid methods have been proposed to merge the computational efficiency of indirect methods with the direct data-to-control philosophy. These approaches typically pre-compute the $\mathbf{g}(k)$ vector before solving the main control problem. In our prior work \cite{mazare2025noise}, we introduced a novel hybrid DPC algorithm that specifically tackled the critical issue of robustness to measurement noise. While other hybrid methods relied on techniques like LQ factorization, our approach, Noise Tolerant DPC (NTDPC), utilized Singular Value Decomposition (SVD). The SVD is uniquely adept at separating the system's core dynamics from stochastic noise within the data Hankel matrix. This is crucial, as measurement noise can obscure the true rank of the system, degrading the estimation of $\mathbf{g}(k)$. Our NTDPC decomposition of $\mathbf{g}(k)$ into its past and future components resulted in a robust and computationally efficient control algorithm for Linear Time-Invariant (LTI) systems.

However, the aforementioned approaches, including NTDPC, are fundamentally built on Willems' lemma, which is valid only for LTI systems. Extending the benefits of DPC to the vast class of nonlinear systems is a critical and highly active area of research. To bridge this gap, kernel-based methods have emerged as a principled and powerful strategy \cite{de2025kernelized, huang2023robust, de2024kernelized}.

Kernel methods handle nonlinearity by implicitly mapping the system data into a high-dimensional (or even infinite-dimensional) reproducing kernel Hilbert space (RKHS) \cite{huang2023robust, de2024kernelized}. In this feature space, complex nonlinear dynamics can often be approximated as linear relationships \cite{huang2023robust}. This "kernel trick" allows the powerful machinery of linear DPC to be applied, effectively linearizing the system dynamics without explicitly constructing the nonlinear model. Recent efforts in this domain, such as Robust Kernelized DeePC (RoKDeePC) \cite{huang2023robust}, have used the representer theorem \cite{scholkopf2001generalized} to implicitly learn the system's nonlinear behavior. Other works have utilized kernelized operator-theoretic approaches to manage large datasets efficiently \cite{de2025kernelized}, or have applied kernels to parameterize the specific nonlinear terms of a velocity model to achieve offset-free control \cite{de2024kernelized}.

While effective, existing kernelized methods often revert to high-dimensional direct DPC formulations, necessitating complex non-convex min-max optimizations to ensure robustness \cite{huang2023robust}. To address this, we extend our hybrid framework by applying SVD factorization directly within the RKHS. This novel approach pre-computes a robust, low-dimensional feature representation that effectively manages nonlinear dynamics and noise, thereby achieving computational efficiency without requiring large-scale online optimization.

\section{Problem Formulation and Kernel Setup}

To extend our hybrid NTDPC framework from the linear domain to nonlinear systems, we leverage the function approximation power of kernel methods. The central idea is to implicitly map the system's trajectory data into a high-dimensional (or even infinite-dimensional) feature space, known as a Reproducing Kernel Hilbert Space (RKHS). We aim to learn a linear predictor in this lifted space that effectively captures the complex nonlinear dynamics \cite{huang2023robust}.

\subsection{Notation and Data Organization}
Consider a discrete-time nonlinear system with inputs $\mathbf{u}(t) \in \mathbb{R}^{n_u}$ and outputs $\mathbf{y}(t) \in \mathbb{R}^{n_y}$. At time step $k$, we define the \textit{past} trajectory vectors $\mathbf{u}_{\text{ini}}(k)$ and $\mathbf{y}_{\text{ini}}(k)$ of length $T_{\text{ini}}$, and the \textit{future} trajectory vectors $\mathbf{u}_N(k)$ and $\mathbf{y}_N(k)$ of length $N$:
\begin{align}
    \mathbf{u}_{\text{ini}}(k) &= \text{col}(\mathbf{u}(k-T_{\text{ini}}), \dots, \mathbf{u}(k-1)) \in \mathbb{R}^{n_u T_{\text{ini}}}, \\
    \mathbf{y}_{\text{ini}}(k) &= \text{col}(\mathbf{y}(k-T_{\text{ini}}), \dots, \mathbf{y}(k-1)) \in \mathbb{R}^{n_y T_{\text{ini}}}, \\
    \mathbf{u}_N(k) &= \text{col}(\mathbf{u}(k), \dots, \mathbf{u}(k+N-1)) \in \mathbb{R}^{n_u N}, \\
    \mathbf{y}_N(k) &= \text{col}(\mathbf{y}(k), \dots, \mathbf{y}(k+N-1)) \in \mathbb{R}^{n_y N}.
\end{align}
To ensure offset-free tracking, we operate on input increments $\Delta \mathbf{u}(t) = \mathbf{u}(t) - \mathbf{u}(t-1)$. We define the combined past data vector $\mathbf{d}_{\text{ini}}(k)$ and the future input increment vector $\mathbf{d}^u_f(k)$ as:
\begin{equation} \label{eq:data_vectors}
    \mathbf{d}_{\text{ini}}(k) = \begin{bmatrix} \Delta \mathbf{u}_{\text{ini}}(k) \\ \mathbf{y}_{\text{ini}}(k) \end{bmatrix}, \quad \mathbf{d}^u_f(k) = \Delta \mathbf{u}_N(k).
\end{equation}

\subsection{Reproducing Kernel Hilbert Spaces (RKHS)}

An RKHS $\mathcal{H}$ over a non-empty set $\mathcal{X}$ is a Hilbert space of functions $f: \mathcal{X} \rightarrow \mathbb{R}$ endowed with a "reproducing kernel" $k: \mathcal{X} \times \mathcal{X} \rightarrow \mathbb{R}$. This kernel is symmetric and positive definite. The defining property of an RKHS is the reproducing property:
\begin{equation}
    \langle f, k(\cdot, \mathbf{x}) \rangle_{\mathcal{H}} = f(\mathbf{x}), \quad \forall f \in \mathcal{H}, \forall \mathbf{x} \in \mathcal{X}.
\end{equation}
This implies that function evaluation is equivalent to an inner product in the feature space.

A cornerstone of kernel methods is the Representer Theorem \cite{scholkopf2001generalized}. It states that the minimizer $f^*$ of a regularized empirical risk function
\begin{equation} \label{eq:risk_min}
    f^* = \arg\min_{f \in \mathcal{H}} \frac{1}{T} \sum_{j=1}^{T} L(y_j, f(\mathbf{x}_j)) + \gamma \|f\|_{\mathcal{H}}^2,
\end{equation}
admits a finite-dimensional representation as a linear combination of kernels centered at the training points $\mathbf{x}_j$:
\begin{equation}
    f^*(\mathbf{x}) = \sum_{j=1}^{T} \alpha_j k(\mathbf{x}_j, \mathbf{x}),
\end{equation}
where $L$ is a loss function, $\gamma > 0$ is a regularization parameter, and $\boldsymbol{\alpha} = [\alpha_1, \dots, \alpha_T]^\top \in \mathbb{R}^T$ are the coefficients to be determined.

\begin{assumption}[Linear Embedding in RKHS] \label{ass:rkhs}
We assume the unknown nonlinear system dynamics can be approximated by a linear map in the RKHS. Specifically, there exists a feature map $\phi: \mathcal{X} \rightarrow \mathcal{H}$ such that the prediction of future outputs can be modeled linearly in the feature space of past data and future inputs.
\end{assumption}

This assumption allows us to generalize the linear NTDPC framework \cite{mazare2025noise} to nonlinear systems. The core NTDPC optimization problem remains a convex Quadratic Program (QP):

\textbf{Problem 1 [NTDPC] \cite{mazare2025noise}}:
At each time step $k$, solve:
\begin{align}
\min_{\Theta} \quad & \|\tilde{\mathbf{y}}_N(k)\|_Q^2 + \|\Delta \mathbf{u}_N(k)\|_R^2  + \|\boldsymbol{\sigma}_y(k)\|_{\Lambda_y}^2 \notag \\
\text{s.t.} \quad & \hat{\mathbf{y}}_N(k) = \mathbf{P}_1 \mathbf{d}_{\text{ini}}(k)  + \mathbf{P}_2 \Delta \mathbf{u}_N(k) + \boldsymbol{\sigma}_y(k), \label{eq:NTDPC_constraint} \\
& \Delta \mathbf{u}_N(k) \in \mathcal{U}, \quad \hat{\mathbf{y}}_N(k) \in \mathcal{Y}, \notag
\end{align}
where $\Theta = \{\Delta \mathbf{u}_N(k), \boldsymbol{\sigma}_y(k)\}$ are the decision variables, $\tilde{\mathbf{y}}_N = \hat{\mathbf{y}}_N - \mathbf{y}_{\text{ref}}$ is the tracking error, and $\mathbf{P}_1, \mathbf{P}_2$ are predictor matrices. In this work, we replace the linear $\mathbf{P}_1, \mathbf{P}_2$ with kernelized predictors $\mathbf{P}_1^{\mathrm{K}}$ and $\mathbf{P}_2^{\mathrm{K}}$ derived from data.

\subsection{Offline Data Collection and Kernel Matrices}

We collect an offline dataset of $T$ trajectories. For the $i$-th trajectory ($i=1,\dots,T$), we store the past data vector $\mathbf{d}_{\text{ini},i}$ and the future input increment vector $\mathbf{d}^u_{f,i}$ as defined in \eqref{eq:data_vectors}. The corresponding future output trajectory is denoted $\mathbf{y}_{f,i} \in \mathbb{R}^{n_y N}$.
We aggregate the output training data into the matrix:
\begin{equation}
\mathbf{Y}_f = \begin{bmatrix} \mathbf{y}_{f,1} & \mathbf{y}_{f,2} & \cdots & \mathbf{y}_{f,T} \end{bmatrix} \in \mathbb{R}^{n_y N \times T}.
\end{equation}

From these trajectory vectors, we construct the kernel (Gram) matrices. The  Past Data Gram Matrix  $\mathbf{K}_{pp} \in \mathbb{R}^{T \times T}$ captures the similarity between past trajectories:
\begin{equation}
[\mathbf{K}_{pp}]_{ij} = k_{\text{ini}}(\mathbf{d}_{\text{ini},i}, \mathbf{d}_{\text{ini},j}). \label{eq:kernel_pp}
\end{equation}
The  Future Input Gram Matrix  $\mathbf{K}_{ff} \in \mathbb{R}^{T \times T}$ captures the similarity between future input sequences:
\begin{equation}
[\mathbf{K}_{ff}]_{ij} = k_{u}(\mathbf{d}^u_{f,i}, \mathbf{d}^u_{f,j}). \label{eq:kernel_ff}
\end{equation}
Here, $k_{\text{ini}}(\cdot, \cdot)$ and $k_{u}(\cdot, \cdot)$ are positive definite kernel functions chosen appropriately for the state and input spaces, respectively (e.g., Gaussian RBF kernels).

\begin{definition}[Persistence of Excitation in RKHS]
The dataset $\{\mathbf{d}_{\text{ini},i}\}_{i=1}^T$ is said to be persistently exciting in the RKHS if the Gram matrix $\mathbf{K}_{pp}$ is full rank, i.e., $\lambda_{\min}(\mathbf{K}_{pp}) > 0$.
\end{definition}

\section{Kernelized Predictor Formulation}

To enable the formulation of the control problem as a Quadratic Program (QP), we require a predictive model that is linear in the decision variables (future input increments) but retains the capacity to capture nonlinear system dynamics from past data. We achieve this by learning a linear operator in the RKHS.

Unlike previous approaches that might treat past and future data streams independently, we formulate a \textit{joint} learning problem to ensure consistency and avoid over-parametrization of the output energy. We define a combined feature vector for the system trajectory $\mathbf{z}(k) = (\mathbf{d}_{\text{ini}}(k), \mathbf{d}^u_f(k))$. We employ a  sum kernel  structure to model the joint similarity:
\begin{equation} \label{eq:joint_kernel}
    k_{\text{joint}}(\mathbf{z}_i, \mathbf{z}_j) = k_{\text{ini}}(\mathbf{d}_{\text{ini},i}, \mathbf{d}_{\text{ini},j}) + k_{u}(\mathbf{d}^u_{f,i}, \mathbf{d}^u_{f,j}).
\end{equation}
This choice implicitly corresponds to a concatenation of feature maps in the RKHS, i.e., $\phi(\mathbf{z}) = [\phi_{\text{ini}}(\mathbf{d}_{\text{ini}})^\top, \phi_{u}(\mathbf{d}^u_f)^\top]^\top$. This structural choice is crucial as it allows us to additively decompose the prediction into a ``past'' component (fixed at time $k$) and a ``future'' component (dependent on the decision variables), while learning them simultaneously against the output data $\mathbf{Y}_f$.

\subsection{Joint Kernel Ridge Regression}

We seek a linear map $\mathcal{W}: \mathcal{H} \rightarrow \mathbb{R}^{n_y N}$ that minimizes the regularized empirical risk over the training dataset:
\begin{equation}
    \min_{\mathcal{W}} \sum_{i=1}^{T} \|\mathbf{y}_{f,i} - \mathcal{W}\phi(\mathbf{z}_i)\|_2^2 + \lambda \|\mathcal{W}\|_F^2,
\end{equation}
where $\lambda > 0$ is a regularization parameter. By the generalized Representer Theorem \cite{scholkopf2001generalized}, the optimal predictor takes the form:
\begin{equation} \label{eq:representer_sol}
    \hat{\mathbf{y}}_N(k) = \mathbf{Y}_f (\mathbf{K}_{\text{joint}} + \lambda \mathbf{I})^{-1} \mathbf{k}_{\text{joint}}(\mathbf{z}(k)),
\end{equation}
where $\mathbf{K}_{\text{joint}} = \mathbf{K}_{pp} + \mathbf{K}_{ff} \in \mathbb{R}^{T \times T}$ is the joint Gram matrix constructed from the offline data sums.
We define the \textit{coefficient matrix} $\mathbf{A} \in \mathbb{R}^{n_y N \times T}$ as:
\begin{equation} \label{eq:coeff_matrix}
    \mathbf{A} = \mathbf{Y}_f (\mathbf{K}_{pp} + \mathbf{K}_{ff} + \lambda \mathbf{I})^{-1}.
\end{equation}
Substituting the sum kernel definition \eqref{eq:joint_kernel} into \eqref{eq:representer_sol}, the prediction naturally separates into two terms sharing the same coefficient matrix $\mathbf{A}$:
\begin{equation}
    \hat{\mathbf{y}}_N(k) = \underbrace{\mathbf{A} \mathbf{k}_{p,\text{ini}}(k)}_{\hat{\mathbf{y}}_N^{\text{past}}(k)} + \underbrace{\mathbf{A} \mathbf{k}_{f,N}(k)}_{\hat{\mathbf{y}}_N^{\text{future}}(k)}.
\end{equation}
Here, $\mathbf{k}_{p,\text{ini}}(k) \in \mathbb{R}^T$ and $\mathbf{k}_{f,N}(k) \in \mathbb{R}^T$ are the online kernel similarity vectors for the past data and future inputs, respectively, as defined in Section II.C.

\begin{remark}
This joint formulation corrects a common pitfall in kernelized control where separate regressions are performed for past and future components on the same target $\mathbf{Y}_f$. By solving for $\mathbf{A}$ using the joint kernel, we ensure that the superposition of past and future effects accurately reconstructs the training outputs without double-counting the signal energy.
\end{remark}

\subsection{Linearization for QP Formulation}

The term $\hat{\mathbf{y}}_N^{\text{past}}(k)$ is fully determined by measured data at time $k$ and acts as a constant offset in the optimization. However, the term $\hat{\mathbf{y}}_N^{\text{future}}(k) = \mathbf{A} \mathbf{k}_{f,N}(\Delta \mathbf{u}_N(k))$ is nonlinear in the decision variable $\Delta \mathbf{u}_N(k)$ (embedded within the kernel function).

To obtain a convex QP, we linearize the future predictor around the equilibrium point $\Delta \mathbf{u}_N = \mathbf{0}$. The first-order Taylor expansion is given by:
\begin{equation}
    \hat{\mathbf{y}}_N^{\text{future}}(k) \approx \hat{\mathbf{y}}_N^{\text{future}}(\mathbf{0}) + \nabla_{\Delta \mathbf{u}_N} \hat{\mathbf{y}}_N^{\text{future}} \Big|_{\mathbf{0}} \Delta \mathbf{u}_N(k).
\end{equation}
Given that we operate on increments, we assume the kernel satisfies $k_u(\mathbf{0}, \mathbf{0}) = c$ (constant) and the system is at a steady state relative to increments, implying the zero-increment prediction offset is absorbed into the bias or zeroed by data centering. The Jacobian of the predictor is:
\begin{equation}
    \mathbf{P}_2^{\mathrm{K}} = \mathbf{A} \left( \frac{\partial \mathbf{k}_{f,N}(\Delta \mathbf{u}_N)}{\partial \Delta \mathbf{u}_N} \bigg|_{\Delta \mathbf{u}_N = \mathbf{0}} \right)^\top.
\end{equation}
For a standard Gaussian RBF kernel with bandwidth $\sigma$, the Jacobian of the kernel vector $\mathbf{J}_{\mathbf{k}} \in \mathbb{R}^{T \times n_u N}$ at zero is:
\begin{equation}
    [\mathbf{J}_{\mathbf{k}}]_{i,:} = -\frac{1}{\sigma^2} k_u(\mathbf{d}^u_{f,i}, \mathbf{0}) (\mathbf{d}^u_{f,i})^\top.
\end{equation}
Thus, the explicit predictor matrices for the QP problem in (Problem 1) are identified as:
\begin{align}
    \mathbf{P}_1^{\mathrm{K}} &= \mathbf{A}, \quad \text{(acting on the kernel feature vector } \mathbf{k}_{p,\text{ini}}), \label{eq:P1_def} \\
    \mathbf{P}_2^{\mathrm{K}} &= \mathbf{A} \mathbf{J}_{\mathbf{k}}. \label{eq:P2_def}
\end{align}
This formulation provides a local linear approximation of the learned nonlinear manifold, valid within a trust region of small input increments.

\begin{assumption}[Stabilizability of the Learned Model] \label{ass:stabilizability}
We assume that the pair $(\mathbf{A}_{lin}, \mathbf{B}_{lin})$ implied by the linearized predictor matrices $(\mathbf{P}_1^{\mathrm{K}}, \mathbf{P}_2^{\mathrm{K}})$ is stabilizable. This ensures that the optimization problem has a feasible solution that converges to the equilibrium.
\end{assumption}

\section{Offset-Free Control Formulation and Analysis}

We now formulate the Kernelized Data-Driven Predictive Control (KDPC) problem. A key innovation of our approach is the specific choice of kernel structure—using a nonlinear kernel for past data complexity and a linear kernel for future input increments. This structure renders the predictive model linear in the decision variables, enabling the control problem to be solved as a strictly convex Quadratic Program (QP) with global optimality guarantees, avoiding the pitfalls of local minima common in nonlinear MPC.

\subsection{Predictor Structure and QP Formulation}

The data-driven predictor derived in Section III, under the choice of a linear kernel for future inputs ($k_u(\mathbf{u}_i, \mathbf{u}_j) = \mathbf{u}_i^\top \mathbf{u}_j$), simplifies to the following semilinear form:
\begin{equation} \label{eq:predictor_structure}
    \hat{\mathbf{y}}_N(k) = \mathbf{P}_1^{\mathrm{K}} \mathbf{k}_{p,\text{ini}}(\mathbf{d}_{\text{ini}}(k)) + \mathbf{P}_2^{\mathrm{K}} \Delta \mathbf{U}_N(k) + \boldsymbol{\sigma}_y,
\end{equation}
where $\mathbf{k}_{p,\text{ini}}(\cdot)$ is the nonlinear feature vector (e.g., RBF) of the past trajectory, and $\mathbf{P}_2^{\mathrm{K}}$ is a \textit{constant} matrix. This significantly improves upon prior works that require real-time Jacobians approximation, as $\mathbf{P}_2^{\mathrm{K}}$ here is exact and pre-computable.

At each time step $k$, we solve the following optimization problem $\mathcal{P}_N(\boldsymbol{\xi}_k)$:
\begin{subequations} \label{eq:KDPC_Problem}
\begin{align}
    \min_{\Delta \mathbf{U}_N, \boldsymbol{\sigma}_y} \quad & J_N(\hat{\mathbf{y}}_N, \Delta \mathbf{U}_N, \boldsymbol{\sigma}_y) \\
    \text{s.t.} \quad & \hat{\mathbf{y}}_N(k) = \mathbf{P}_1^{\mathrm{K}} \mathbf{k}_{p,\text{ini}}(k) + \mathbf{P}_2^{\mathrm{K}} \Delta \mathbf{U}_N + \boldsymbol{\sigma}_y, \label{eq:model_const} \\
    & \Delta \mathbf{u}_{k+j|k} \in \mathcal{U}, \quad \forall j \in [0, N-1], \\
    & \hat{\mathbf{y}}_{k+j|k} \in \mathcal{Y} \oplus \mathcal{B}(\boldsymbol{\sigma}_y), \quad \forall j \in [0, N-1], \\
    & \hat{\mathbf{y}}_{k+N|k} \in \mathcal{Y}_f. \label{eq:terminal_const}
\end{align}
\end{subequations}
The cost function $J_N$ is defined as:
\begin{align}
    J_N =& \sum_{j=0}^{N-1} \Big( \|\hat{\mathbf{y}}_{k+j|k} - \mathbf{y}_{\text{ref}}\|_Q^2 + \|\Delta \mathbf{u}_{k+j|k}\|_R^2 \Big) \nonumber \\& + \|\hat{\mathbf{y}}_{N} - \mathbf{y}_{\text{ref}}\|_P^2 + \lambda_{\sigma} \|\boldsymbol{\sigma}_y\|^2
\end{align}
where $Q, R \succ 0$ are stage costs, $P \succ 0$ is the terminal cost, and $\lambda_{\sigma}$ penalizes the slack variable. The constraint \eqref{eq:terminal_const} enforces the terminal state to belong to a robust invariant set $\mathcal{Y}_f$, ensuring recursive feasibility.

\begin{remark}[Convexity and Real-Time Feasibility]
A distinct advantage of the proposed formulation is its computational tractability. Unlike recent data-driven velocity-form approaches (e.g., [11]), which formulate the control problem as a Nonlinear Program (NLP) solved via iterative Sequential Quadratic Programming (SQP), our specific choice of a linear kernel for future inputs renders problem $\mathcal{P}_N$ a strictly convex Quadratic Program (QP). This guarantees the existence of a unique global optimum and allows for the use of efficient, deterministic solvers suitable for high-frequency real-time control, avoiding the potential for local minima or varying convergence times associated with SQP methods.
\end{remark}

\begin{algorithm}[t]
\caption{Robust Offset-Free KDPC }
\label{alg:KDPC}
\begin{algorithmic}[1]
\REQUIRE Offline data $\{\mathbf{d}_{\text{ini},i}, \mathbf{d}^u_{f,i}, \mathbf{y}_{f,i}\}_{i=1}^T$; Weights $Q, R, P$.
\STATE \textbf{Offline Training:}
\STATE Compute Gram matrices: $\mathbf{K}_{pp}$ (RBF on past data) and $\mathbf{K}_{ff} = (\mathbf{H}^u_{fut})^\top \mathbf{H}^u_{fut}$ (Linear on inputs).
\STATE Compute coefficient matrix $\mathbf{A} = \mathbf{Y}_f (\mathbf{K}_{pp} + \mathbf{K}_{ff} + \lambda \mathbf{I})^{-1}$.
\STATE Extract predictors: $\mathbf{P}_1^{\mathrm{K}} = \mathbf{A}$ and $\mathbf{P}_2^{\mathrm{K}} = \mathbf{A} \mathbf{H}^u_{fut}$.
\STATE Compute Terminal Set $\mathcal{Y}_f$ based on the implied linear model pair $(\mathbf{A}_{lin}, \mathbf{B}_{lin})$.
\STATE \textbf{Online Control Loop:}
\FOR{$k = 0, 1, 2, \dots$}
    \STATE Measure $\mathbf{y}(k)$ and update history $\mathbf{d}_{\text{ini}}(k)$.
    \STATE Calculate kernel vector $\mathbf{k}_{p,\text{ini}}(k) = k_{rbf}(\mathbf{H}_{past}, \mathbf{d}_{\text{ini}}(k))$.
    \STATE Solve QP \eqref{eq:KDPC_Problem} to obtain $\Delta \mathbf{u}^*_{k|k}$.
    \STATE Apply $\mathbf{u}(k) = \mathbf{u}(k-1) + \Delta \mathbf{u}^*_{k|k}$.
\ENDFOR
\end{algorithmic}
\end{algorithm}

\subsection{Stability Analysis}

We analyze the stability of the closed-loop system. Acknowledging that kernel methods provide an approximation of the true dynamics, we establish Input-to-State Stability (ISS) rather than asymptotic stability.

\begin{assumption}\label{ass:error}[Bounded The mismatch between the true nonlinear system dynamics $f_{true}(\cdot)$ and the linearized kernel predictor is bounded on the compact operating set:
\begin{equation}
    \| f_{true}(\xi, \Delta u) - f_{ker}(\xi, \Delta u) \| \leq \epsilon.
\end{equation}
This uniform error bound $\epsilon$ captures both the kernel regression residual and the linearization error. The existence of such finite-data error bounds for kernel-based control surrogates is rigorously established in \cite{bold2025kernel} and \cite{schimperna2025data}, justifying the validity of this assumption.
\end{assumption}

\begin{assumption}[Terminal Ingredients] \label{ass:terminal}
There exists a terminal control law $\kappa_f(\mathbf{y}) = K_f \mathbf{y}$ and an invariant set $\mathcal{Y}_f$ such that $\forall \mathbf{y} \in \mathcal{Y}_f$, the system remains in $\mathcal{Y}_f$ and the cost decreases: $V_f(f(\mathbf{y}, \kappa_f(\mathbf{y}))) - V_f(\mathbf{y}) \leq - \ell(\mathbf{y}, \kappa_f(\mathbf{y}))$.
\end{assumption}

\begin{theorem}[Recursive Feasibility]
\label{thm:rec_feas}
Let Assumptions \ref{ass:error} (Bounded Approximation Error) and \ref{ass:terminal} (Robust Terminal Ingredients) hold. If the optimization problem $\mathcal{P}_N(\boldsymbol{\xi}_k)$ is feasible at time $k$, then it is feasible at time $k+1$ for any measured state $\boldsymbol{\xi}_{k+1}$ satisfying the error bound $\|\boldsymbol{\xi}_{k+1} - \hat{\boldsymbol{\xi}}_{1|k}\| \leq \epsilon$.
\end{theorem}

\begin{proof}
Assume $\mathcal{P}_N(\boldsymbol{\xi}_k)$ is feasible and let $\Delta \mathbf{U}^*(k) = \{\Delta \mathbf{u}^*_{0|k}, \dots, \Delta \mathbf{u}^*_{N-1|k}\}$ be the optimal input sequence, generating the optimal predicted trajectory $\mathbf{Y}^*(k) = \{\hat{\mathbf{y}}^*_{1|k}, \dots, \hat{\mathbf{y}}^*_{N|k}\}$. By the terminal constraint, $\hat{\mathbf{y}}^*_{N|k} \in \mathcal{Y}_f$.

We construct a feasible candidate solution for time $k+1$ as follows:

\textbf{1. Candidate Input Sequence:}
We shift the optimal input sequence and append the terminal control law:
\[
\Delta \tilde{\mathbf{U}}(k+1) := \{ \Delta \mathbf{u}^*_{1|k}, \dots, \Delta \mathbf{u}^*_{N-1|k}, \kappa_f(\hat{\mathbf{y}}^*_{N|k}) \}.
\]
Since $\Delta \mathbf{u}^*_{i|k} \in \mathcal{U}$ for $i=1,\dots,N-1$ (from time $k$) and $\kappa_f(\hat{\mathbf{y}}^*_{N|k}) \in \mathcal{U}$ (by the control invariance of $\mathcal{Y}_f$ in Assumption \ref{ass:terminal}), the candidate input sequence satisfies the hard input constraints: $\Delta \tilde{\mathbf{U}}(k+1) \in \mathcal{U}^N$.

\textbf{2. Prediction Consistency via Slack Variables:}
The prediction equality constraint at time $k+1$ is given by:
\[
\hat{\mathbf{y}}_N(k+1) = \mathbf{P}_1^{\mathrm{K}} \mathbf{k}_{p,\text{ini}}(\boldsymbol{\xi}_{k+1}) + \mathbf{P}_2^{\mathrm{K}} \Delta \tilde{\mathbf{U}}(k+1) + \tilde{\boldsymbol{\sigma}}_y.
\]
The measured state $\boldsymbol{\xi}_{k+1}$ differs from the predicted predecessor $\hat{\boldsymbol{\xi}}_{1|k}$ by the bounded error $\epsilon$ (Assumption \ref{ass:error}). Since $\tilde{\boldsymbol{\sigma}}_y$ is a free decision variable in the optimization problem $\mathcal{P}_N$, there exists a candidate slack $\tilde{\boldsymbol{\sigma}}_y$ that absorbs this mismatch and the model linearization error perfectly. Specifically, we choose $\tilde{\boldsymbol{\sigma}}_y$ such that the equality holds. Consequently, the soft output constraints $\hat{\mathbf{y}}_{j|k+1} \in \mathcal{Y} \oplus \mathcal{B}(\tilde{\boldsymbol{\sigma}}_y)$ are satisfied by definition.

\textbf{3. Terminal Constraint Satisfaction:}
We must ensure the terminal predicted state $\hat{\mathbf{y}}^{cand}_{N|k+1}$ lies within $\mathcal{Y}_f$.
The candidate trajectory applies the optimal inputs shifted by one step. The state at the end of the horizon, $\hat{\mathbf{y}}^{cand}_{N|k+1}$, corresponds to applying the terminal law $\kappa_f$ to the state $\hat{\mathbf{y}}^*_{N|k}$ subject to the propagated error from $\boldsymbol{\xi}_{k+1}$.
However, Assumption \ref{ass:terminal} posits that $\mathcal{Y}_f$ is a \textit{Robust} Positively Invariant (RPI) set for the closed-loop system under disturbances bounded by $\epsilon$. Since $\hat{\mathbf{y}}^*_{N|k} \in \mathcal{Y}_f$ and the accumulated error remains within the robust margin handled by $\mathcal{Y}_f$ and the slack variables, the terminal state remains feasible: $\hat{\mathbf{y}}^{cand}_{N|k+1} \in \mathcal{Y}_f$.

Thus, a feasible solution $\{\Delta \tilde{\mathbf{U}}(k+1), \tilde{\boldsymbol{\sigma}}_y\}$ exists, proving recursive feasibility.
\end{proof}

\begin{remark}[Feasibility Strategies: Slacks vs. Tightening]
We highlight that the use of slack variables $\boldsymbol{\sigma}_y$ is a design choice to ensure recursive feasibility without the conservatism of \textit{constraint tightening}. Recent rigorous data-driven frameworks (e.g., \cite{schimperna2025data}) typically employ robust constraint tightening (Tube MPC) to guarantee that the system state strictly satisfies hard constraints despite prediction errors. While theoretically elegant, calculating the required Robust Positively Invariant (RPI) sets for kernel-based models can be computationally prohibitive. Our soft-constraint formulation offers a practical trade-off: it guarantees solver feasibility at every step (Theorem \ref{thm:rec_feas}) while relying on the ISS property (Theorem \ref{thm:ISS}) to bound the constraint violations by the learning error $\epsilon$.
\end{remark}

\begin{theorem}[Input-to-State Stability]
\label{thm:ISS}
Let Assumptions \ref{ass:error} (Bounded Kernel Error) and \ref{ass:terminal} (Robust Terminal Ingredients) hold. The closed-loop system generated by the proposed KDPC is Input-to-State Stable (ISS) with respect to the kernel approximation error $\epsilon$. Specifically, the system state $\boldsymbol{\xi}_k$ converges to a robust positively invariant set $\Omega(\epsilon)$ defined by
\[
\Omega(\epsilon) := \{ \boldsymbol{\xi} \in \mathbb{R}^{n_\xi} \mid \alpha(\|\boldsymbol{\xi}\|) \le \gamma(\epsilon) \},
\]
where $\alpha(\cdot) \in \mathcal{K}_\infty$ is bounded by the stage cost and $\gamma(\cdot) \in \mathcal{K}$ depends on the Lipschitz constant of the value function.
\end{theorem}

\begin{proof}
Consider the optimal value function $V(\boldsymbol{\xi}_k) := J^*_N(\boldsymbol{\xi}_k)$ as the ISS-Lyapunov candidate. We examine the evolution of $V(\cdot)$ along the closed-loop trajectory $\boldsymbol{\xi}_{k+1} = f_{true}(\boldsymbol{\xi}_k, \Delta \mathbf{u}^*_{0|k})$.

Let $\Delta \mathbf{U}^*(k) = \{ \Delta \mathbf{u}^*_{0|k}, \dots, \Delta \mathbf{u}^*_{N-1|k} \}$ be the optimal input sequence at time $k$, generating the predicted state trajectory $\{ \hat{\boldsymbol{\xi}}_{0|k}, \dots, \hat{\boldsymbol{\xi}}_{N|k} \}$.
By Theorem 1 (Recursive Feasibility), a feasible candidate sequence $\Delta \tilde{\mathbf{U}}(k+1)$ exists, constructed by shifting $\Delta \mathbf{U}^*(k)$ and appending the terminal law $\kappa_f(\hat{\boldsymbol{\xi}}_{N|k})$.
The cost of this candidate sequence, denoted $\tilde{J}_N(\hat{\boldsymbol{\xi}}_{1|k})$, satisfies the standard nominal descent property derived from the terminal cost condition (Assumption \ref{ass:terminal}):
\begin{equation} \label{eq:nominal_descent}
\tilde{J}_N(\hat{\boldsymbol{\xi}}_{1|k}) - V(\boldsymbol{\xi}_k) \le -\ell(\boldsymbol{\xi}_k, \Delta \mathbf{u}^*_{0|k}) \le -\alpha(\|\boldsymbol{\xi}_k\|),
\end{equation}
where $\alpha(\|\cdot\|) := \lambda_{\min}(Q) \|\cdot\|^2 \in \mathcal{K}_\infty$.

The optimization problem $\mathcal{P}_N$ is a strictly convex Quadratic Program (QP) defined on compact constraints. As established in multi-parametric programming theory, the value function $V(\cdot)$ of such a QP is piecewise quadratic and continuous on its domain. Since the feasible set is compact, $V(\cdot)$ is locally Lipschitz continuous. Thus, there exists a constant $L_V > 0$ such that for any two feasible states $\boldsymbol{\xi}, \boldsymbol{\xi}'$:
\[
|V(\boldsymbol{\xi}) - V(\boldsymbol{\xi}')| \le L_V \| \boldsymbol{\xi} - \boldsymbol{\xi}' \|.
\]

At time $k+1$, the measured state is $\boldsymbol{\xi}_{k+1}$. Due to the kernel approximation error (Assumption \ref{ass:error}), $\boldsymbol{\xi}_{k+1}$ differs from the predicted state $\hat{\boldsymbol{\xi}}_{1|k}$ by at most $\epsilon$:
\[
\| \boldsymbol{\xi}_{k+1} - \hat{\boldsymbol{\xi}}_{1|k} \| \le \epsilon.
\]
By optimality, $V(\boldsymbol{\xi}_{k+1}) \le J_N(\boldsymbol{\xi}_{k+1}, \Delta \tilde{\mathbf{U}}(k+1))$. Utilizing the Lipschitz property and the triangle inequality:
\begin{align*}
V(\boldsymbol{\xi}_{k+1}) - V(\boldsymbol{\xi}_k) &= V(\boldsymbol{\xi}_{k+1}) - \tilde{J}_N(\hat{\boldsymbol{\xi}}_{1|k}) + \tilde{J}_N(\hat{\boldsymbol{\xi}}_{1|k}) - V(\boldsymbol{\xi}_k) \\
&\le |V(\boldsymbol{\xi}_{k+1}) - \tilde{J}_N(\hat{\boldsymbol{\xi}}_{1|k})| - \alpha(\|\boldsymbol{\xi}_k\|) \\
&\le L_V \| \boldsymbol{\xi}_{k+1} - \hat{\boldsymbol{\xi}}_{1|k} \| - \alpha(\|\boldsymbol{\xi}_k\|) \\
&\le - \alpha(\|\boldsymbol{\xi}_k\|) + L_V \epsilon.
\end{align*}
This inequality satisfies the definition of Input-to-State Stability. The state $\boldsymbol{\xi}_k$ converges to the level set where $\alpha(\|\boldsymbol{\xi}\|) \le L_V \epsilon$, defining the robust invariant set $\Omega(\epsilon)$ with $\gamma(\epsilon) = L_V \epsilon$. This confirms that the asymptotic tracking error is bounded by the quality of the kernel learning $\epsilon$.
\end{proof}

\begin{remark}[Stability under Approximation Errors]
Recent theoretical works (e.g., \cite{bold2025kernel, schimperna2025data}) have established asymptotic stability for data-driven MPC by requiring the approximation error to vanish proportionally to the state norm. In contrast, our approach accepts a non-vanishing uniform error $\epsilon$—a consequence of our efficient analytical linearization—and establishes Input-to-State Stability (ISS). This guarantees convergence to a robust invariant set $\Omega(\epsilon)$ rather than the origin, offering a practical trade-off for real-time computational efficiency.
\end{remark}

\section{Illustrative Example}

To validate the proposed KDPC scheme, we apply it to the discrete-time Van der Pol oscillator ($\mu = 1.0$, $T_s = 0.05$ s) subject to matched input disturbance $d(k)$:
\begin{align} \label{eq:vdp_discrete}
\begin{bmatrix}
x_1(k+1) \\
x_2(k+1)
\end{bmatrix}
=&
\begin{bmatrix}
1 & T_s \\
-T_s & 1
\end{bmatrix}
\begin{bmatrix}
x_1(k) \\
x_2(k)
\end{bmatrix}
+
\begin{bmatrix}
0 \\
T_s
\end{bmatrix}
(u(k) + d(k))
 \nonumber \\&+
\begin{bmatrix}
0 \\
T_s \mu (1 - x_1^2(k)) x_2(k)
\end{bmatrix},
\end{align}
with output $y(k) = x_1(k)$. We benchmark the purely data-driven KDPC against a standard model-based NMPC. The KDPC is trained on $T_{\text{offline}}=500$ samples collected using random excitation $u \in [-10, 10]$. It employs a hybrid kernel structure (RBF with $\gamma_p = 0.001$ for past data; linear for future inputs) and the proposed feedback bias correction. In contrast, the NMPC utilizes the exact analytic model \eqref{eq:vdp_discrete} but lacks specific knowledge of $d(k)$ or explicit integral action. Both controllers share the prediction horizon $N=10$, initialization $T_{\text{ini}}=5$, and weights $Q=10.0, R=0.05$.

\subsection{Results and Discussion}

\begin{figure}[!t]
\centerline{\includegraphics[width=\columnwidth]{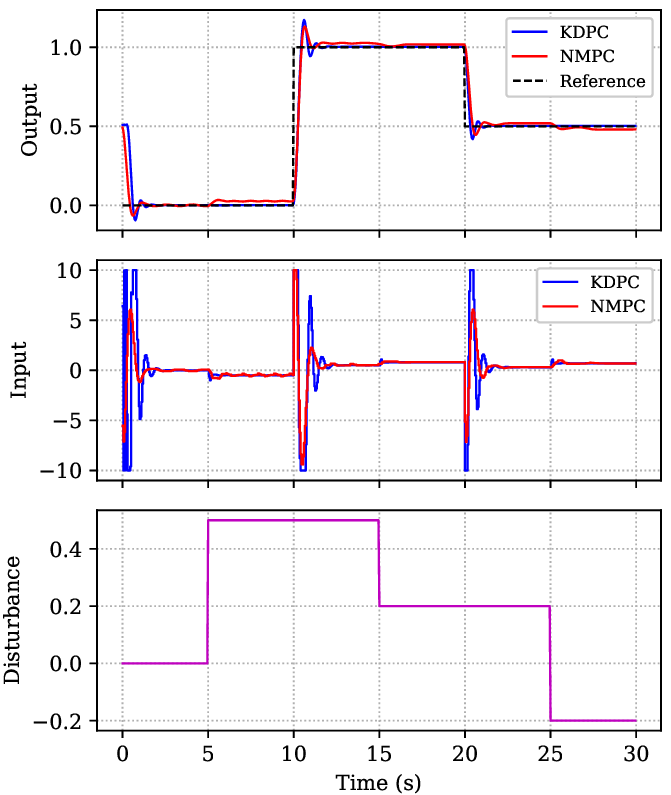}}
\caption{Comparative simulation results for the Van der Pol oscillator. \textbf{Top:} Output tracking performance. The proposed KDPC (blue) achieves zero steady-state error despite the unmeasured input disturbance (magenta, bottom panel), whereas the standard NMPC (red) exhibits a persistent offset. \textbf{Middle:} Control input trajectories. \textbf{Bottom:} Profile of the unmeasured input disturbance $d(k)$.}
\label{fig:U1}
\end{figure}

The closed-loop performance is evaluated under a challenging scenario involving a piecewise constant reference $y_{\text{ref}}$ and a concurrent, unmeasured piecewise constant input disturbance $d(k)$, as shown in Fig.~\ref{fig:U1}. The results highlight the ``shining property'' of the proposed KDPC: its inherent capability for robust offset-free tracking. When a large unmeasured input disturbance ($d=0.5$) is introduced at $t=5$ s, the standard NMPC fails to compensate, settling at a persistent steady-state error despite possessing a perfect system model. This failure arises because the standard formulation lacks the internal model principle mechanism required for rejecting constant input disturbances. In sharp contrast, the KDPC rapidly detects the mismatch through its feedback bias correction term. This mechanism functions as an integral action, effectively estimating the disturbance and shifting the predicted trajectory to drive the tracking error back to zero, validating the Input-to-State Stability properties established in Theorem 2.

Beyond disturbance rejection, the hybrid kernel architecture demonstrates significant robustness during reference steps (e.g., $t=10$s and $t=20$s) where the Van der Pol oscillator exhibits strong nonlinearity. The KDPC tracks these transitions as accurately as the model-based benchmark, confirming that the learned RBF kernel successfully captures the underlying nonlinear manifold without explicit modeling. Furthermore, the control input trajectories (middle panel) reveal that the KDPC generates a smooth, realizable signal that naturally respects constraints ($u \in [-10, 10]$). Unlike data-driven methods prone to high-frequency chattering, the KDPC maintains a consistent input profile comparable to the NMPC, attributed to the Tikhonov regularization which prevents overfitting to measurement noise.

\subsection{Computational Complexity Analysis}

The computational complexity of the proposed kernelized offset-free KDPC framework can be systematically decomposed into offline and online phases. Offline, the dominant operations include the inversion of the $T \times T$ kernel matrices $\mathbf{K}_{pp}$ and $\mathbf{K}_{ff}$, incurring $\mathcal{O}(T^3)$ time complexity each, followed by matrix multiplications for constructing the predictor matrices $\mathbf{P}_1^{\mathrm{K}}$ and $\mathbf{P}_2^{\mathrm{K}}$, which require $\mathcal{O}(T^2 n_y N)$ operations.

Online, the computational advantage of the proposed method becomes evident. By restricting the kernel formulation to be linear in future inputs (Section III.B), the resulting optimization problem is a strictly convex Quadratic Program (QP). Computing the kernel similarity vector $\mathbf{k}_{p,\mathrm{ini}}(k)$ scales as $\mathcal{O}(T \cdot \dim(\mathbf{d}_{\mathrm{ini},i}))$. The QP solver typically exhibits $\mathcal{O}((n_u N + n_y N)^3)$ complexity per iteration for interior-point methods.
In our simulations, the proposed KDPC demonstrated significantly faster solve times compared to the NMPC, which requires solving a non-convex Nonlinear Programming (NLP) problem at every step. This makes the KDPC highly suitable for real-time control of nonlinear systems with fast sampling rates.

\section{Conclusion}

This paper presented a novel offset-free Kernelized Data-Driven Predictive Control (KDPC) scheme for nonlinear systems, utilizing kernel methods to learn unknown dynamics directly from data. By integrating a hybrid kernel architecture with a feedback bias correction mechanism, the proposed approach guarantees robust reference tracking without prior model knowledge. Numerical validation on a Van der Pol oscillator demonstrated that KDPC achieves superior performance compared to standard model-based NMPC, specifically by eliminating steady-state errors in the presence of unmeasured input disturbances where NMPC failed. These results confirm KDPC as a powerful, data-efficient strategy for achieving robust, high-performance control of unknown nonlinear systems.

\end{document}